\documentclass[letterpaper,prl,twocolumn,superscriptaddress]{revtex4}

\usepackage{graphicx,bm}

\begin{document}

\title{Conductance through a single impurity in the metallic zigzag carbon nanotube}

\author{Po-Yao Chang}
\affiliation{ Department of Physics, National Tsing-Hua University,
Hsinchu 30013, Taiwan}

\author{Hsiu-Hau Lin}
\affiliation{ Department of Physics, National Tsing-Hua University,
Hsinchu 30013, Taiwan}
\affiliation{Physics Division, National
Center for Theoretical Sciences, Hsinchu 30013, Taiwan}
\date{\today}
\begin{abstract}
We investigate transport through a single impurity in metallic zigzag carbon nanotube and find the conductance sensitively depends on the impurity strength and the bias voltage. It is rather interesting that interplay between the current-carrying scattering state and evanescent modes leads to rich phenomena including resonant backward scattering, perfect tunneling and charge accumulations. In addition, we also find a dual relation between the backscattered conductance and the charge accumulation. At the end, relevance to the experiments is discussed.
\end{abstract}
\maketitle


Single-wall carbon nanotubes (CNTs)\cite{Charlier07} are graphene sheet\cite{Geim07} in roll-up geometry with large aspect ratio, making them naturally-made quantum wires with interesting transport properties. Since the conducting channels in metallic CNT are limited, defects and impurities\cite{Kim03,Ishigami04} are important for realistic considerations and may play a significant role in transport at nanoscale. One of the recent breakthroughs is the technique of covalent attachments\cite{Brett07} to CNTs, making the fabrication of single-molecule junctions possible. It is rather remarkable that one can use the conductance to monitor and manipulate the covalent attachments to the targeted CNT. On the theoretical side, vacancy defect or doped boron/nitrogen in CNTs\cite{Choi00,Neophytou07} show conductance suppression due to enhanced backward scattering.

Inspired by these experimental and theoretical studies, we investigate the differential conductance through a single impurity in the metallic zigzag CNT. Note that there are many types of defects including pentagon-heptagon pairs, vacancies and chemisorption adatoms\cite{Hashimoto04}. For quantitative comparison with experiments, the first-principles approaches are more appropriate. On the other hand, to reveal the close connections between the conductance, the transmission matrix and the current-induced charge accumulation, a phenomenological approach may prove to be useful. As long as the impurity potential is short-ranged, the detail profile is not crucially important and can be captured by the strength of the pseudo-potential, which is taken to be a delta-function for simplicity. These simplifications allow us to compute the scattering states in the metallic CNT and construct the transmission matrix exactly. Making use of the Landauer-B\"{u}ttiker formula, we can compute the differential conductance at different energies.

At first glance, this may seem to be a simple calculation leading to trivial results -- one expects the conductance is close to perfect for weak impurity strength, while it should get strongly suppressed for a strong impurity potential. This naive expectation is not correct since impurities in honeycomb lattice cause strong resonant states and non-trival evanescent modes. As a result, the conductance sensitively depends on the strength of the impurity potential and the bias voltage. In the absence of the impurity, the two conducting channels in metallic zigzag CNT deliver the perfect conductance $G = 2 G_0$, where $G_0 = 2e^2/h$ is the quantum conductance. In some energy regime, we found the backward scattering is greatly enhanced so that one of the conducting channel is perfectly blocked, leading to resonant backward scattering. On the other hand, in some other energy regime, the perfect conductance is recovered, signaling the conspiracy from the evanescent modes to make the impurity potential transparent. In the following, we would elaborate on the detail derivations that provides a firm ground for the rich physics of the conductance in CNTs.


\begin{figure}
\centering
\includegraphics[width=7cm]{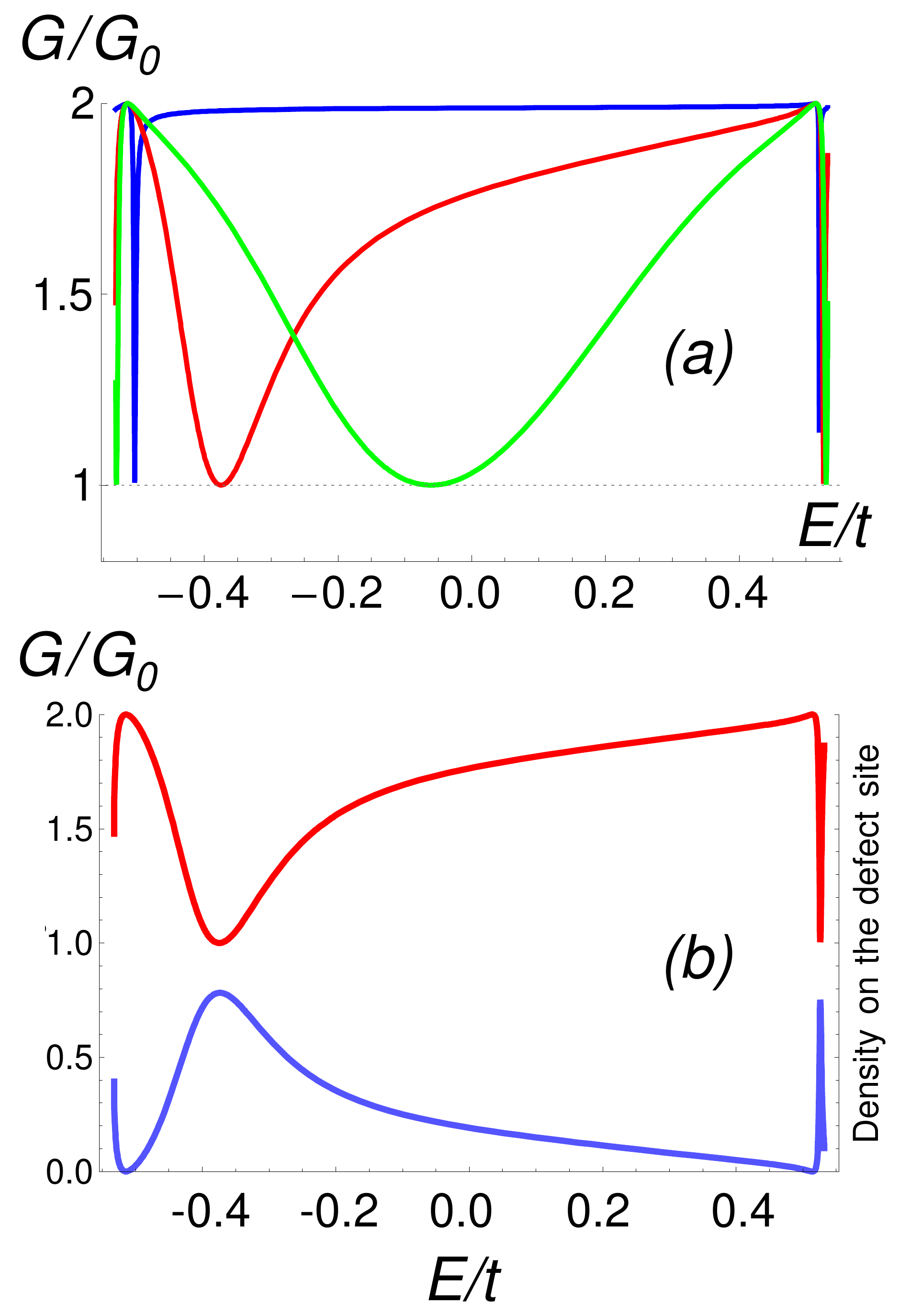}
\caption{\label{Fig1} (Color online) (a) Conductance versus energy for $(9,0)$-zigzag CNT with different impurity strength $V_0/t = 1, 5, 50$ shown in blue (dark), red (gray) and green (light gray) respectively. (b) The energy-dependent conductance (upper curve) and the charge density on the impurity site (lower curve) with $V_0/t = 5$ are plotted together for comparison.}
\end{figure}

In the following calculations, we concentrate on the metallic zigzag CNT that is well described by the tight-binding model,
\begin{equation}\label{e1}
\textit{H}=-t\sum_{\langle i,j\rangle}\left(c^\dagger_i c_j+c^\dagger_j c_i \right)
+ V_0 c^\dag_0 c_0,
\end{equation}
where $t$ is the hopping amplitude between nearest neighbors and $V_0$ is the strength of the impurity potential located at the origin. Since no spin-flip scattering is present, the spin just gives rise to a factor of two and is suppressed in the above expression.

The above Hamiltonian can be solved by the generalized Bloch theorem\cite{Lin05} where the momentum is allowed to be complex due to the presence of the impurity potential. In the following, we sketch how this is done in metallic zigzag CNT. In the absence of the impurity, the transverse momentum is a good quantum number, $k_m=2\pi m/N$ with $m=1,2,...,N$ and $N$ is the number of unit cells around the CNT. Performing a partial Fourier transformation for the wave function in the transverse direction, $\Phi(x,y)=\sum_{k} \Phi_{k}(x)e^{iky}$, the CNT is decomposed into $N$ decoupled effective 1D chains with definite transverse momentum $k_m$. In general, the wave function after partial Fourier transformation take the form, $\Phi_k(x) \sim \phi_k z^x$. For $|z|=1$, these solutions correspond to the familiar plane waves. However, in the presence of the impurity potential, $|z| \neq 1$ solutions are also allowed and need to be taken into consideration. For given energy $\epsilon$ and momentum $k_m$, $z$ satisfies the characteristic equation,
\begin{eqnarray}
t t_m(z+\frac{1}{z})+(t^2+t^2_m-\epsilon^2)=0,
\end{eqnarray} 
where $t_m=2t \cos(k_m/2)$. By forming appropriate linear combinations of all possible solutions of $z$, the scattering states can be constructed analytically and thus the transmission matrix is obtained. Since there are only two scattering channels in a metallic zigzag CNT, the $4 \times 4$ $S$-matrix is
\begin{eqnarray}
  S(\epsilon)=\left[
        \begin{array}{cc}
          r(\epsilon) & t'(\epsilon) \\
          t(\epsilon) & r'(\epsilon) \\
        \end{array}
      \right],
\end{eqnarray}
where $r,r',t,t'$ are $2 \times 2$ matrices. Finally, the Landauer-B\"{u}ttiker formula gives the conductance as tracing over the square of the transmission matrix,
\begin{equation}
G(\epsilon)= G_0 {\rm Tr} \left[t^\dagger(\epsilon) t(\epsilon) \right].
\end{equation}
The transmission matrix can be derived analytically, involving complicated expressions which can be evaluated by numerical means.

\begin{figure}
\centering
\includegraphics[width=6.5cm]{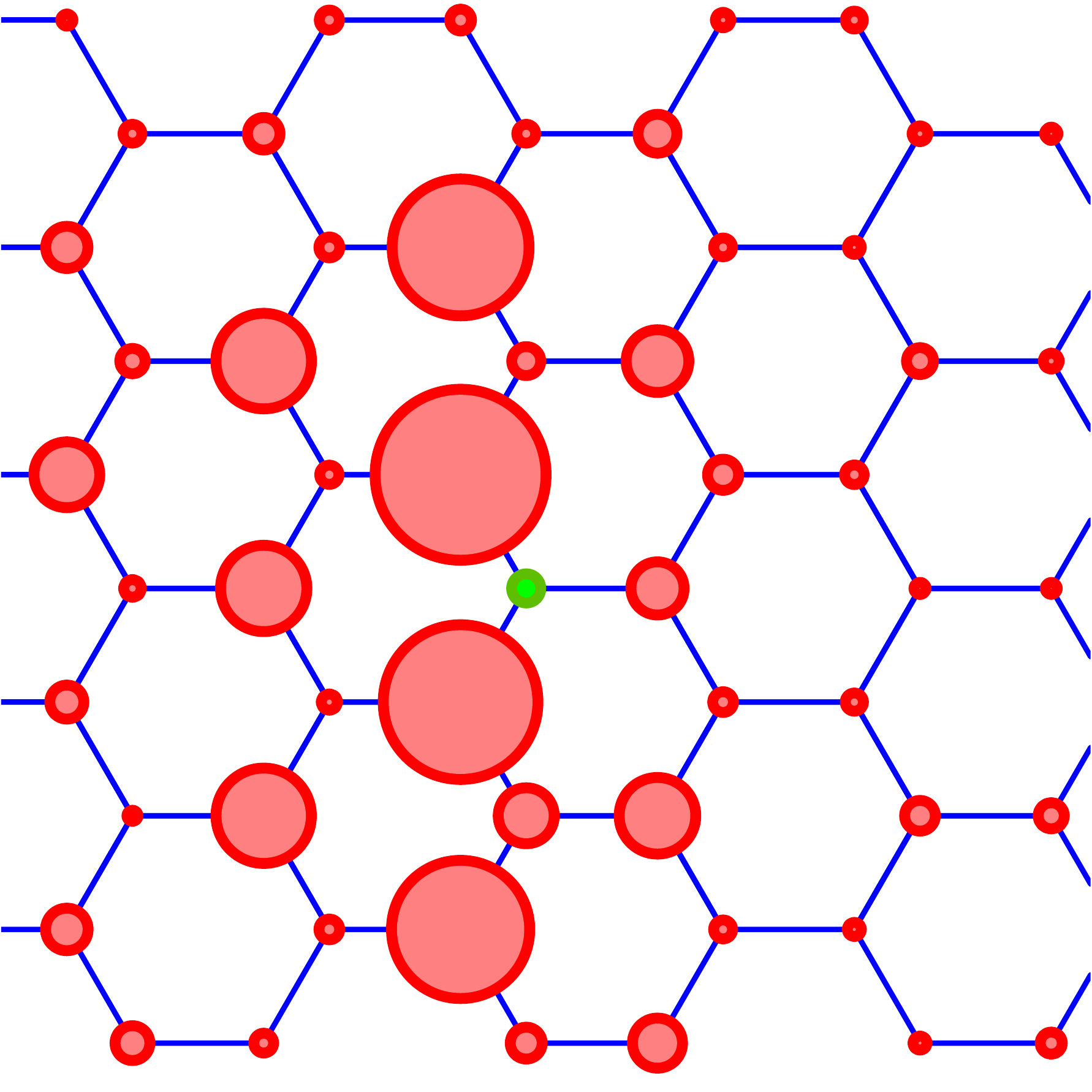}
\caption{\label{Fig2} (Color online) Charge accumulation at perfect blocking $G=G_0$ in the $(9,0)$-zigzag CNT with $V_0/t = 1$. The radius of the shaded area represents the charge density with the defect site labeled in green (lighter gray).}
\end{figure}


In Fig. \ref{Fig1}, the numerical results are shown. First of all, since the impurity potential only exists at the origin, one can always form an appropriate linear combination of the two conducting channels so that one of them remains intact from scattering. This ensures that the conductance is always large than $G_0$. For weak impurity potential $V_0/t = 1$, the conductance is close to the perfect value $G \sim 2 G_0$ at most energies. However, there exist two sharp dips down to exactly $G = G_0$ where one of the channel is completely blocked. We found the perfect blocking is directly tied up with the resonant backward scattering with the evanescent modes. Though the conductance merely comes from the scattering states, the evanescent modes are also induced during the impurity scattering process. In Fig.~\ref{Fig2}, the charge accumulation at perfect blocking energy is shown with relatively weak impurity strength $V_0/t=1$. It is remarkable that such a large charge accumulation can be created by a weak impurity potential on one site. By increasing the impurity strength, the resonant backward scattering is broadened but the perfect blocking $G=G_0$ remains. Finally, strong impurity strength brings the system to the unitary limit where the particle-hole symmetry is restored.

Another interesting feature is the tunneling conductance becomes perfect at some energies. This may not look very surprising for weak impurity strength. However, even for $V_0/t =50$, the strong impurity potential is still transparent at some particular energies as shown in Fig.~\ref{Fig1}. How can the scattering states ignore the strong impurity scattering and maintain the conducting channel perfect? To understand this phenomena, we plot the conductance and the particle density on the impurity site together in Fig.~\ref{Fig1}(b) for comparison. When the density on the impurity site is large, the conductance has a dip, while when the density is zero, the conductance is perfect. This correlation can be understood from the Lippmann-Schwinger equation -- with vanishing density on the defect site, the scattering state remains the same as the unperturbed wave function and thus the impurity potential looks transparent.

\begin{figure}
\centering
\includegraphics[width=7cm]{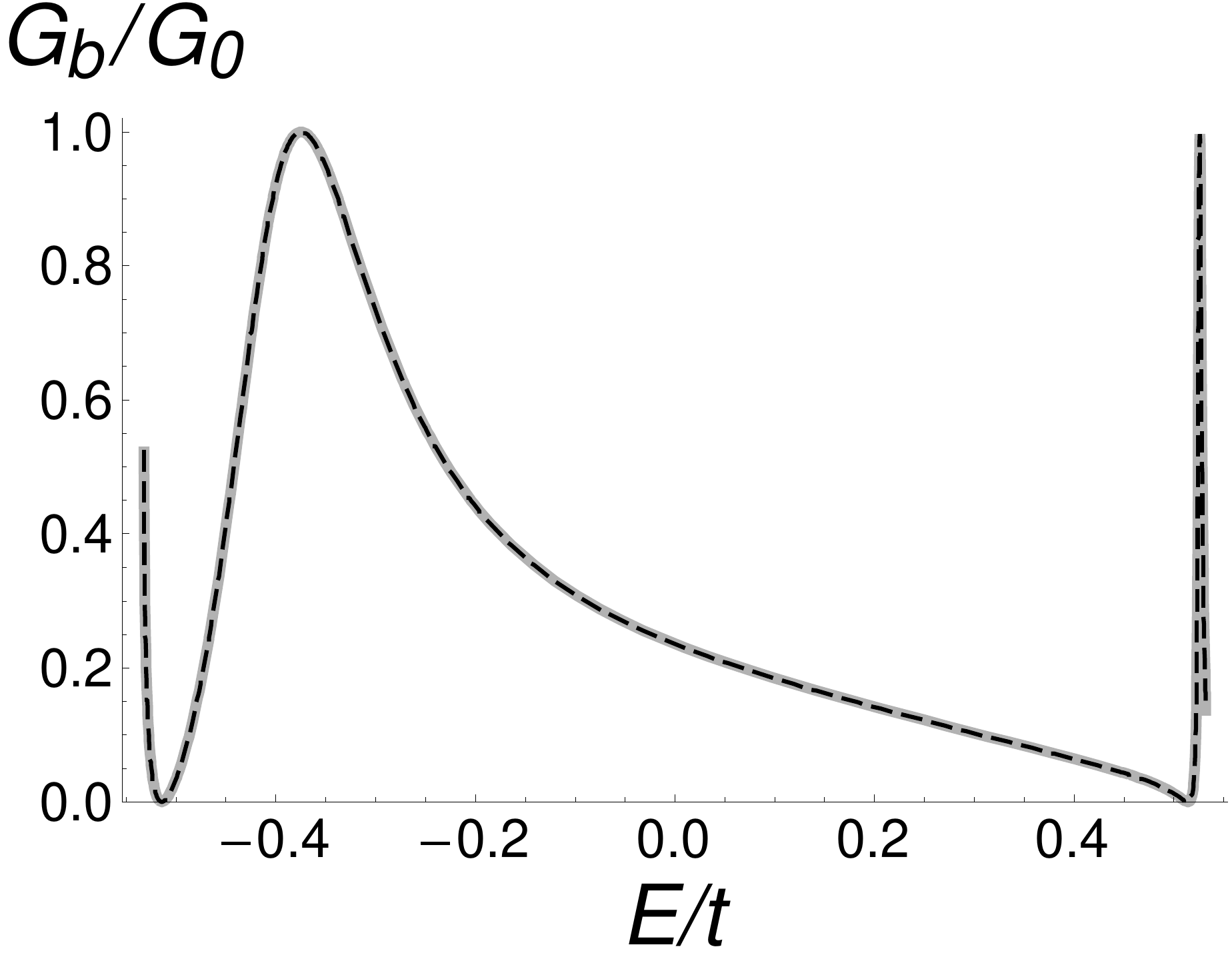}
\caption{\label{Fig3} Dimensionless backscattered conductance $g_b \equiv G_b/G_0$ and $\sin^2 \eta_t$ versus energy plotted in solid and dashed lines. The identical match of these two curves agrees with the analytic proof shown in the text.}
\end{figure}


The single-impurity scattering consider here is the dual junction for the tunneling setup through a disconnected quantum dot. In the tunneling junction, the Friedel sum rule relates the conductance to the phase of the transmission matrix $\eta_t$ and the  $G= G_0 \sin^2 \eta_t = G_0 \sin^2 (\pi \Delta N)$, where $\eta_t$ is the phase of $\det t(\epsilon)$ and $\Delta N$ is the change of the particle number. It is rather interesting that a dual relation for backscattered conductance is found,
\begin{eqnarray}
G_b \equiv 2G_0 - G = G_0 \sin^2 \eta_t,
\end{eqnarray}
for the single-impurity scattering as well. We checked the relation is valid for our numerical results as shown in Fig.~\ref{Fig3}. According to Friedel sum rule, $\Delta N =
\eta_t/\pi$. Thus, we come to a useful relation $G_b=G_0\sin^2 (\pi\Delta N)$ that establishes the connection between the backscattered conductance and the charge accumulation.

To show the relation between $G_b$ and $\eta_t$, we start from the Dyson equation for the Green's function,
\begin{eqnarray}
G_{xx'}(\omega) = G^0_{xx'}(\omega) + \sum_{x_1,x_2}
G^0_{xx_1}(\omega) T_{x_1x_2} G^0_{x_2x'}(\omega),
\end{eqnarray}
where $G$ and $G^0$ are the Green's functions with and without the impurity potential and $T_{xx'}$ is the $T$-matrix in coordinate representation. The non-interacting Green's function can be computed by contour integral,
$G^0_{xx'}(\omega) = -i \pi \rho(\omega) e^{ik_\omega|x-x'|}$,
where $k_\omega= \epsilon^{-1}(\omega+\mu)$ is the magnitude of the momentum and $\rho(\omega)$ is the density of states. Note that the tunneling amplitude $t(\omega) = G_{xx'}(\omega)/G^0_{xx'}(\omega)$ is the ratio of the Green's functions in the asymptotic limit, $x \to \infty$ and $x' \to -\infty$, where the Dyson equation takes the simple form after some algebra,
$
G_{xx'}(\omega) = G^0_{xx'}(\omega) [1-i\pi \rho(\omega) T(\omega)].
$
By comparison, we obtain the key relation between the tunneling amplitude and the $T-$matrix, $t(\omega) = 1-i\pi \rho(\omega) T(\omega)$.
Making use of the relation ${\cal S} = 1-2\pi i \rho T$ and the unitarity ${\cal S} = e^{2i\delta}$, the tunneling amplitude is $t(\omega) =  \cos \delta\: e^{i\delta}$. It is clear that $G_b = G_0 (1-\cos^2 \delta)$ and $\eta_t = \delta$ so that the relation is established.


Our findings here echo with the recent experiment\cite{Bockrath01} that the defect in CNT can turn opaque in the Coulomb blockade regime or transparent at other bias voltages. Furthermore, there exist controlled methods to fabricate the defects\cite{Krasheninnikov07} systematically, or manipulate the effective impurity strength\cite{Parka07} by a voltage pulse from the tip of the atomic force microscope on CNT. Therefore, the resonant backward scattering, perfect tunneling and charge accumulations discussed here are relevant and can be observed in experiments on transport through the single defect in CNT.

We acknowledge supports from the National Science Council in Taiwan through grants NSC-96-2112-M-007-004 and NSC-97-2112-M-007-022-MY3. Financial supports and friendly environment provided by the National Center for Theoretical Sciences in Taiwan are also greatly appreciated.


\end{document}